\documentclass[sn-mathphys-num]{sn-jnl}

\usepackage{graphicx}%
\usepackage{multirow}%
\usepackage{amsmath,amssymb,amsfonts}%
\usepackage{amsthm}%
\usepackage{mathrsfs}%
\usepackage[title]{appendix}%
\usepackage{xcolor}%
\usepackage{textcomp}%
\usepackage{svg}%
\usepackage{manyfoot}%
\usepackage{booktabs}%
\usepackage{algorithm}%
\usepackage{algorithmicx}%
\usepackage{algpseudocode}%
\usepackage{listings}%
\DeclareUnicodeCharacter{03B1}{\ensuremath{\alpha}}
\DeclareUnicodeCharacter{03B2}{\ensuremath{\beta}}
\DeclareUnicodeCharacter{03B3}{\ensuremath{\gamma}}
\usepackage{booktabs} 
\usepackage{graphicx} 

\theoremstyle{thmstyleone}%
\theoremstyle{thmstyletwo}%

\theoremstyle{thmstylethree}%

\begin{document}

\title[Article Title]{Cross-modal Medical Image Generation Based on Pyramid Convolutional Attention Network}

\author[1]{\fnm{Fuyou} \sur{Mao}}\email{maoeyu@csu.edu.cn}
\author[1]{\fnm{Lixin} \sur{Lin}}\email{llx@csu.edu.cn}
\author[3]{\fnm{Ming} \sur{Jiang}}\email{mjiang@guet.edu.cn}
\author[1]{\fnm{Dong} \sur{Dai}}\email{daidon07@163.com}
\author[2]{\fnm{Chao} \sur{Yang}}\email{234712253@csu.edu.cn}
\author*[1]{\fnm{Hao} \sur{Zhang}}\email{hao@csu.edu.cn}
\author*[1]{\fnm{Yan} \sur{Tang}}\email{tangyan@csu.edu.cn}

\affil*[1]{\orgdiv{School of Electronic Information}, \orgname{Central South University}, \orgaddress{\city{Changsha}, \country{China}}}
\affil[2]{\orgdiv{School of Computer Science and Engineering}, \orgname{Central South University}, \orgaddress{\city{Changsha}, \country{China}}}
\affil[3]{\orgdiv{School of Computer Science and Information Security}, \orgname{Guilin University of Electronic Technology}, \orgaddress{\city{Guilin}, \postcode{541000}, \country{China}}}


\abstract{The integration of multimodal medical imaging can provide complementary and comprehensive information for the diagnosis of Alzheimer's disease (AD). However, in clinical practice, since positron emission tomography (PET) is often missing, multimodal images might be incomplete. To address this problem, we propose a method that can efficiently utilize structural magnetic resonance imaging (sMRI) image information to generate high-quality PET images. Our generation model efficiently utilizes pyramid convolution combined with channel attention mechanism to extract multi-scale local features in sMRI, and injects global correlation information into these features using self-attention mechanism to ensure the restoration of the generated PET image on local texture and global structure. Additionally, we introduce additional loss functions to guide the generation model in producing higher-quality PET images. Through experiments conducted on publicly available ADNI databases, the generated images outperform previous research methods in various performance indicators (average absolute error: 0.0194, peak signal-to-noise ratio: 29.65, structural similarity: 0.9486) and are close to real images. In promoting AD diagnosis, the generated images combined with their corresponding sMRI also showed excellent performance in AD diagnosis tasks (classification accuracy: 94.21 \%), and outperformed previous research methods of the same type. The experimental results demonstrate that our method outperforms other competing methods in quantitative metrics, qualitative visualization, and evaluation criteria.}





\maketitle

\section{Introduction}\label{sec1}

Alzheimer's disease (AD) is a long-term neuro-degenerative condition that affects thinking, behavior, and memory. Early identification is extremely important because the disease is progressive, meaning that symptoms gradually get worse over time. However, the country’s diagnosis and treatment rate for AD remains low, with few medical specialists and minimal public awareness\cite{bi2020computer}.

With the advancement of three-dimensional (3D) imaging techniques, medical imaging modalities such as structural magnetic resonance imaging (sMRI)\cite{folego2020alzheimer}and positron emission tomography (PET)\cite{liu2020multi}have become crucial tools for evaluating brain pathology in patients. These medical imaging modalities provide detailed views of the brain structure and function in AD. Recently, accumulating evidence suggests the in-tegration of multimodal medical imaging can provide complementary and comprehen-sive information for the diagnosis of AD\cite{rallabandi2023deep,qiu2022multimodal,shukla2023alzheimer}. However, in clinical practice, the appli-cation of PET imaging is often limited due to factors such as high cost and radiation exposure. This limitation results in incomplete multimodal data, where the availability of PET images may be scarce or insufficient. For example, in the ADNI database, alt-hough all subjects have sMRI data, only approximately half of them underwent PET scans. To address this issue, a common strategy is to estimate the missing data of sub-jects using the average or median features of other subjects with complete data or even using random values\cite{77777donders2006gentle}. However, this approach introduces additional noise and is only applicable to manually extracted features. Another more intuitive strategy is to directly perform cross-modal image translation, which involves generating the missing PET images based on the available sMRI data.

In recent years, deep learning methods have gained significant attention for their ability to generate high-quality images in medical image modality conversion. Li et al\cite{li2014deep}. firstly proposed a 3D Convolutional Neural Networks (3D-CNN) for the transformation from sMRI to PET.  By seizing the nonlinear relationship between these two modalities, they successfully achieved cross-modal conversion from sMRI to PET. However,  relying solely on simple convolution operations makes it challenging to establish the complex nonlinear relationship between sMRI to PET. Sikka et al\cite{sikka2018mri}. firstly proposed  3D-UNet model to capture the non-local relationship and non-linear correlation between these two modalities. The encoder consists of 3D convolution op-eration based on the residual learning concept, where the decoder includes downsam-pling operations with multi-scale convolution operations. Skip connections are used to link these two aspects to enhance the fusion of deep dimensional information. Ronneberger proposed 3D-cGAN method with two generators and discriminators to obtain sMRI-to-PET relationship. However, there is still significant domain gap between sMRI and PET modalities, which leads to suboptimal conversion effects. Gao et al\cite{gao2019deep}. introduced a deep residual Inception encoder-decoder network (RIED-Net) to learn the nonlinear mapping relationship between the input sMRI and the target PET images. Hu et al\cite{hu2019cross}. proposed an effective U-Net architecture with the adversarial training strategy to synthesizing PET datas from their corresponding MRI, and using other popular normalization methods rather than using BN as default to solve the problem of rapid decline in BN performance. However, these CNN-based models often employ L1 or L2 distance as the loss function, both of which produce average results by calculating the average difference between pixel values in the images. This average effect may lead to the blurring or loss of some detailed information during the image generation process. Moreover, due to the locality of CNN, these fully CNN-based methods also face challenges in generating high-quality images with fine-grained textures\cite{isola2017image}.Thus, different from the existing methods, we further considered using pyramid convolution to fully utilize multi-scale local features from sMRI and using a self-attention mechanism to enhance the global representation of feature information. Additionally, a joint loss function was used to guide the generator in producing more realistic PET images.

\section{Related Work}\label{sec2}

In recent years, the deep learning method represented by Generative Adversarial Nets (GAN)\cite{goodfellow2020generative} has achieved remarkable success in various fields for its versatility in image generation. GANs are also widely used in the field of medical image analysis, including denoising\cite{yang2018low} , reconstruction\cite{wolterink2017deep}, segmentation\cite{jiang2018tumor}, etc.

The ability of GANs to generate diverse and realistic image samples makes them particularly attractive to the medical field, which lacks large datasets. Several variants of GANs have been proposed to deal with cross-modal generation or data enhancement in the medical field. Some classical GAN models that have shown good performance in natural image generation tasks, such as conditional GAN (cGAN)\cite{mirza2014conditional} and CycleGAN\cite{zhu2017unpaired}. Hu et al\cite{hu2021bidirectional}. presented a novel 3D end-to-end network, called bidirec-tional mapping GAN (BMGAN) to greatly improve the quality of synthesized images. Lin et al\cite{lin2021bidirectional}. imputed the missing data through the 3D Reversible Generative Adversarial Network (RevGAN) which contains information about AD-related diseases to assist in disease diagnosis. Furthermore, in order to improve the efficiency of generating PET images, Sikka et al\cite{sikka2021mri}.utilized both global and local contextual information from the sMRI image in their GAN architecture (GLA-GAN). Gao et al\cite{gao2021task}. proposed a hybrid deep learning framework that contains a task-induced pyramid and attention generative adversarial network (TPA-GAN) for multimodal image imputation and classification, which greatly improved the ability to classify final diseases. Pan et al\cite{pan2018synthesizing}. proposed a 3D CycleGAN and introduced additional joint loss functions to generate higher-quality PET images in their subsequent work. The studies mentioned above show that GAN is an effective method for data expansion and simulation in segmentation or classification tasks. However, there is still much room to improve the performance in many medical image synthesis tasks. How to balance the quality and efficiency of image generation is a problem that needs to be continuously optimized in medical image synthesis tasks.

Based on the above comparison of literature, we propose a generation model based on pyramid convolution and attention mechanism, which will be used in a more effective way for the synthesis of brain MRI to PET. The main contributions of this study are as follows: First, pyramid convolution was employed to extract multi-scale local features from the sMRI and the channel attention mechanism was introduced to address the issue of information redundancy in pyramid convolution. This helps the model focus on more critical local features during training. Second, we utilized a self-attention mechanism to enhance the global representation of the feature information, ensuring the integrity of the generated images in terms of global structure. Third, we introduced a joint loss function to guide the generator in producing more realistic PET images. The model was extensively evaluated on the publicly available ADNI dataset, and the generated PET images exhibited favorable performance in various evaluation metrics.

\section{Method}\label{sec3}

\subsection{Overview of the Proposed Method}\label{subsec2}

\begin{figure}
    \centering
    \includegraphics[width=1.0\linewidth]{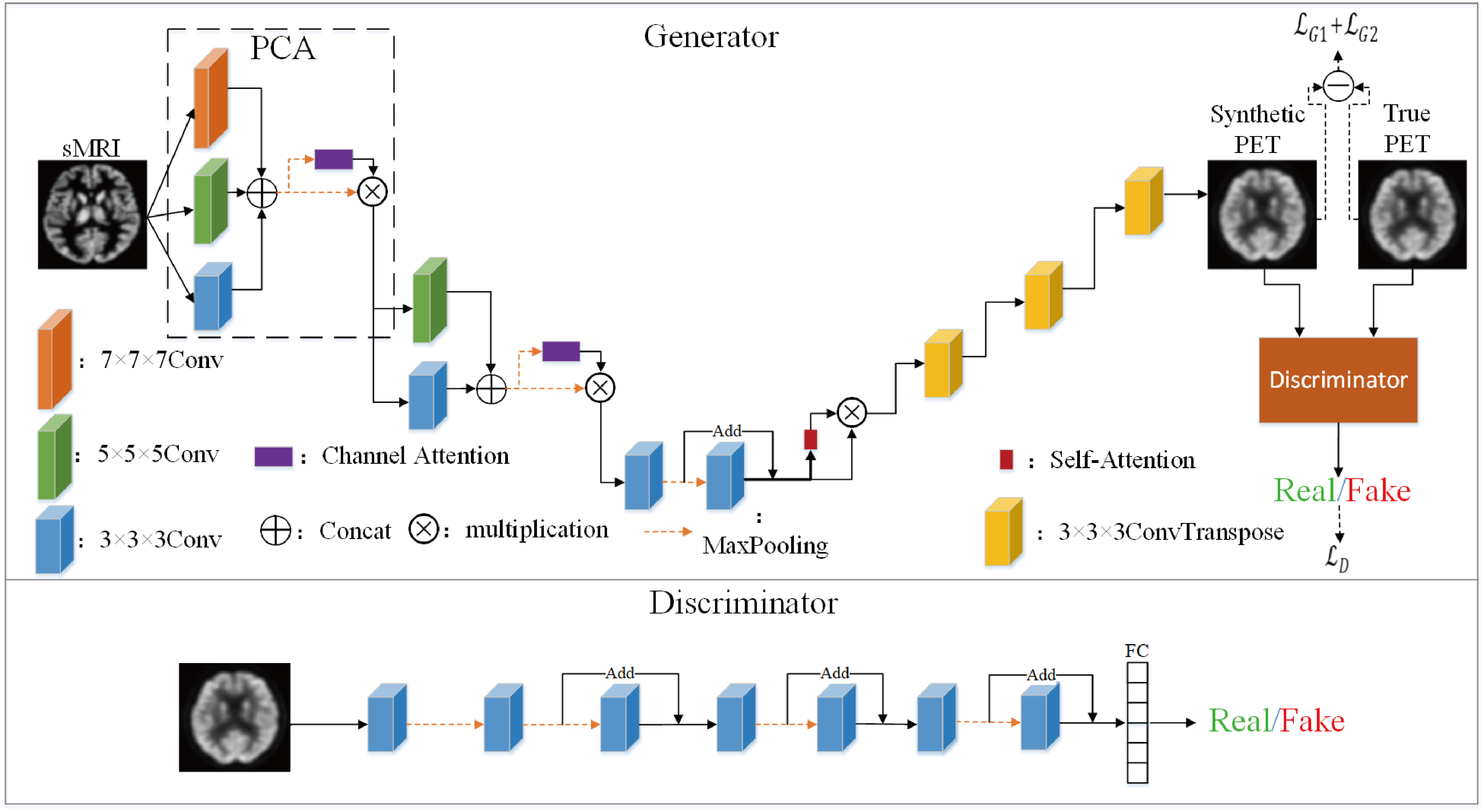}
    \caption{The overall framework of the model}
    \label{fig:Fig1}
\end{figure}

We mainly investigated the issue of missing multimodal image data in clinical practice. We proposed a GAN model for generating missing PET images based on the correlation between PET and sMRI. The GAN consists of a generator and a discriminator. The goal of the generator is to make the missing data as close as possible to real data, while the discriminator learns features to distinguish the input image from the real image. The training process stops when the discriminator fails to recognize the input image as fake. The cross modal medical image generation (sMRI to PET) frame-work proposed in this article is shown in Fig.\ref{fig:Fig1} The details are presented in the following subsections.

\subsection{Generator based on Pyramid Convolutional Attention Network}\label{subsec2}

To generate more realistic PET images, the feature information in sMRI should be fully exploited. Traditional CNNs have a fixed kernel size, which limits their ability to capture features at different scales due to their limited receptive field. Secondly, conventional GAN models do not consider long-range correlations between features, which leads to missing global structures in the generated images. To address the above issues, we propose a generation model based on pyramid convolution, channel attention mechanism, and self-attention mechanism (PCSA-GAN) for the generation of missing PET (see Fig.\ref{fig:Fig1} Generator part).

We constructed a generator model based on the conditional GAN (cGAN) concept and the U-Net structure, which includes two parts: contraction (feature extraction) and reconstruction (image generation). To capture different levels of detailed features from sMRI, we applied two pyramid convolution and channel attention (PCCA) blocks in the contraction part. In the first PCCA block, convolutional kernels of sizes 7×7×7, 5×5×5 , and 3×3×3 were used to extract multi-scale features from the input image, with 4, 8, and 12 kernels per size, respectively. Small convolutional kernels can capture finer details, while large convolutional kernels focus on broader information, improving the abstraction level of features and further enhancing the model's expressive power. Therefore, the operation of pyramid convolution can be defined as:

\begin{align}
  & {{F}_{i}}X=Conv(X,{{k}_{i}},{{C}_{i}}) \\ 
 & F=Concat({{F}_{1}},{{F}_{2}},...,{{F}_{n}})
\end{align}
Where x is the input image or feature map for pyramid convolution; Conv3D rep-resents the 3D convolution operation with kernel size \(k×k×k\) and n filters. \(F1, F2, ..., Fn\) are the features extracted from different scales of convolution operations, and these features are concatenated to obtain the final feature map F.

\begin{figure}
    \centering
    \includegraphics[width=0.8\linewidth]{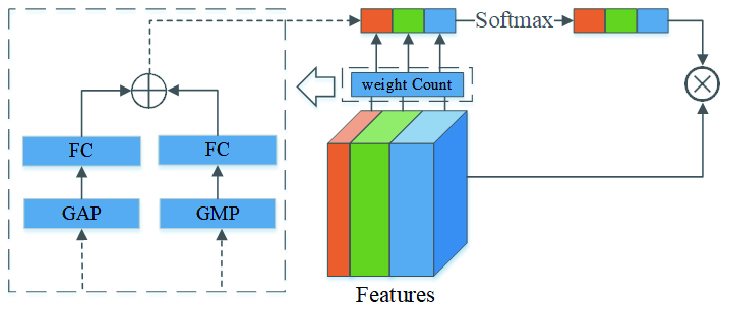}
    \caption{Channel attention mechanism}
    \label{fig:Fig2}
\end{figure}

However, due to the convolution kernels of different scales overlapping with the input image,  some features extracted at one scale may overlap with those extracted at another scale, resulting in redundant feature information. Therefore, the channel attention (CA) mechanism can be used to alleviate this issue. The CA mechanism is a technique used to enhance the ability of CNNs to process feature maps. Its core idea is to enhance the critical information in the features by adaptively learning the weights of each channel. By weighting different feature channels, the model can focus more on important features. The calculation and weighting operation of the CA mechanism are shown in Fig.\ref{fig:Fig2}.

We group the features extracted by convolutions with the same kernel size as one set. Then, we perform global average pooling (GAP) and global max pooling (GMP) operations on each set of features separately, and obtain the weight coefficients of the features through a fully connected layer. The weight coefficients of the two pooling operations are added and activated by the Softmax function to obtain the final weights of each feature in each set. Finally, multiply the weights with their corresponding features, we obtain the weighted features. The formula for channel attention calculation is shown below:

\begin{equation}
CA(X)=\sigma ({{W}_{2}}\delta ({{W}_{1}}GAP(X))+{{W}_{4}}\delta ({{W}_{3}}GMP(X))).\label{eq1}
\end{equation}

The input feature is represented by \(x\) , and GAP and GMP represent global average pooling and global max pooling, respectively. By combining these two pooling operations, more comprehensive feature information can be obtained, and better weighting can be performed. \(W_1\), \(W_2\), \(W_3\) and \(W_4\) represent fully connected (FC) layers,  represents the ReLU activation function, and   represents the Sigmoid activation function. Therefore, the weight vector of the multi-scale channel attention can be represented as follows:

\begin{align}
  & at{{t}_{i}}=Soft\max (CA({{F}_{i}})) \\ 
 & {{Y}_{i}}={{F}_{i}}\odot at{{t}_{i}}
\end{align}

Where, \(atti_i\) represents the channel attention weights for the i-th group of features, and multiplying it with the corresponding group feature \(F_i\) yields the weighted feature \(Y_i\). Finally, concatenating these weighted features together results in the output feature of the PCA layer:

\begin{align}
{{F}_{Out}}=Concat({{Y}_{1}},{{Y}_{2}},...,{{Y}_{n}})
\end{align}

As the feature scale decreases, the number and size of convolution kernels used in the pyramid convolution decrease accordingly. Therefore, in the second PCCA module, only 36 and 60 convolutions are used with kernel sizes of 5×5×5 and 3×3×3, respectively, for multi-scale feature extraction. In the last stage of feature extraction, considering that the feature scale is already small, only 192 convolutions with a scale of 3×3×3 are used for further feature extraction, and the stride of the convolution kernel is set to 2. All convolution operations are followed by ReLU activation function for linear rectification, and max pooling is used for feature dimensionality reduction. Finally, the model enhances feature representation through a residual connection module to obtain the features used for image generation. 

The expansion part of the generator is responsible for generating high-resolution image details from low-resolution feature maps. The high-level abstract features extracted by the contraction part have already integrated the multi-scale local and detail information of the image and have good representational power. However, some global information of the input image may not have been captured, and this information is helpful for maintaining the integrity of the global structure of the generated image. Therefore, self-attention (SA) mechanism can be used to learn the correlations between different position features, and the weights of different position features can be weighted and summed to better represent the low-level features. By applying SA, information loss can be reduced, and the entire network can more effectively utilize the information from the input data, resulting in more realistic image generation. Therefore, SA is used to learn global correlation information among local features, and the features are reconstructed through three 3×3×3 deconvolutional layers with ReLU activation. The changes in feature size and channels during the reconstruction process correspond to those in the feature extraction stage. The linear interpolation algorithm is used as the interpolation algorithm in the image reconstruction process to better reconstruct the details in PET images.

\subsection{Discriminator}\label{subsec2}

The discriminator evaluates the images generated by the generator and provides gradient signals to update the generator's parameters based on the difference between the generated images and the real images. Typically, the discriminator's structure should not be too complex, otherwise the model is prone to overfitting during training, which can prevent the generator from learning effective content and result in distorted generated images. Therefore, four convolutional and pooling operations with a size of 3×3×3 are used to extract effective information from the image, and the number of feature channels after each convolutional layer is 24, 48, 96, and 192, respectively. In the last three convolutional-pooling layers, residual connections are added to enhance the representation of image features, help alleviate the gradient vanishing problem caused by the depth of the network, and accelerate model convergence. Finally, the learned feature information is input into a fully connected (FC) layer to judge the authenticity of the image.

\subsection{Joint Loss Function}\label{subsec2}

To further improve the quality of the generated PET images, we constructed a joint loss function to optimize the network parameters.

\subsubsection{Adversarial loss}\label{subsubsec2}

The adversarial loss function in classical GAN is trained by minimizing the difference between real and generated images, which helps the discriminator to better distinguish between real and generated images. This adversarial training process gradually improves the ability of the discriminator to recognize real and generated images, while also helping the generator to generate more realistic and more in-line with the real distribution of images. The formula for adversarial loss is as follows:

\begin{equation}
    \mathcal{L}_{G1} = \mathbb{E}_{x \sim P_{sMRI}(x), y \sim P_{PET}(y)} \left[ |y - G(x)| \right]
\end{equation}

Where D and G represent the discriminator and the generator, respectively. Ac-cording to this formula, the generator's loss function is implicit, that is, there is no explicit loss function that measures the difference between the generated image and the real image. The quality of the generated image is improved by optimizing the performance of the discriminator. However, due to the complexity of medical images, using only adversarial loss may not be enough to generate high-quality images with the same characteristics as real PET images. Therefore, in this chapter, an additional term is added to the loss function to penalize the loss of detail and structural features, and to improve the quality of the generated images, as described below.

\subsubsection{voxel-wise reconstruction loss (L1)}\label{subsubsec2}

L1 is a standard loss function that can be used to quantify the difference between the generated PET image and the real PET image. The calculation of this loss function is obtained by calculating the difference between the two images at each voxel (3D pixel), and its calculation formula is as follows:

\begin{align}
{{\mathcal{L}}_{G1}}={{E}_{x\sim{{P}_{sMRI(x)}},y\sim{{P}_{PET}}(y)}}[|y-G(x)|]
\end{align}

Where, \(x\) is the input image sMRI, \(y\) is its corresponding ground truth PET, and G is the generator. Minimizing this difference can make the generated image closer to the ground truth image.

\subsubsection{Multi-scale structural similarity index (MS-SSIM) loss}\label{subsubsec2}

In addition to voxel-level loss, we also want to focus on the structural similarity of the images during the image generation process, in order to make the generated PET images more structurally similar to the real PET images. SSIM is another metric used to evaluate image quality, and it has been widely used in image processing algorithms. The calculation formula for SSIM under single scale is as follows:

\begin{align}
[SSIM(x,y)={{[l(x,y)]}^{\alpha }}{{[c(x,y)]}^{\beta }}{{[s(x,y)]}^{\gamma }}\
\end{align}

Where, \(l(x,y)\), \(c(x,y)\), and \(s(x,y)\) are measures of brightness, contrast, and structural similarity of the images, and \(\alpha\), \(\beta\) , and \(\gamma \)   are weighting parameters for each measure. Since SSIM is differentiable, it can be used as a loss function to guide the sensitivity of generated images to structural changes. However, SSIM can only focus on and compare local details in the image and cannot well consider the overall structural features of the image, i.e., lacks perception of large-scale structural changes. Therefore, we adopt the multi-scale version of SSIM (MS-SSIM) as a loss function. To calculate MS-SSIM at M scales, the image needs to be downsampled to 1/2 of its original size and the following formula is used for calculation:

\begin{align}
[MS-SSIM(x,y)={{[l(x,y)]}^{{{\alpha }_{M}}}}\prod\limits_{j=1}^{M}{{{[{{C}_{j}}(x,y)]}^{{{\beta }_{j}}}}{{[{{S}_{j}}(x,y)]}^{{{\gamma }_{j}}}}}\
\end{align}

Where \(α_M\), \(β_j\), and \(γ_j\) are the weight parameters for the luminance \(l(x,y)\), contrast \((x,y)\), and structural similarity measures \(s(x,y) \)at different scales. This chapter selects a scale of \(M=3\) and, following the choice in , sets \(\alpha_{M} = \beta_{j} = \gamma_{j}\) for any scale \(j\), and sets \(\beta_{1} = 0.0448\), \(\beta_{2} = 0.2856\), and  \(\beta_{3} = 0.3001\). This method ensures the structural consistency of the generated images at multiple resolutions, thereby improving the visual effect of the generated images. Since a higher MS-SSIM value indicates a smaller difference between the generated image and the real image, this loss is defined as:

\begin{equation}
    \mathcal{L}_{G2} = \mathbb{E}_{x \sim P_{sMRI(x)}, y \sim P_{PET}(y)} \left[ 1 - MS - SSIM(x, y) \right]
\end{equation}

The overall model uses the following joint loss function for joint optimization to perform cross-modal generation from sMRI to PET.

\begin{equation}
\mathcal{L} = \alpha \mathcal{L}_{D} + \beta \mathcal{L}_{G1} + \gamma \mathcal{L}_{G2}
\end{equation}

Where, \(\alpha\), \(\beta\) and \(\gamma\) are manually set hyperparameters that balance the importance of different loss terms.

\section{Experiments}\label{sec4}
\subsection{Datasets and Processing}\label{subsec1}

In this experiment, we used the open-access sMRI and PET datasets from the Alzheimer's Disease Neuroimaging Initiative (ADNI) database . ADNI is multicenter research to search for clinical, imaging, genetic, and biochemical biomarkers for the discovery of AD. We used data from all subjects who simultaneously had multimodality (MRI and PET), totaling 356 groups which consist of 117 AD subjects and 239 cognitively normal (CN) subjects. The male-to-female ratio is 188/168. The age of the subjects ranges from 56 to 96. In addition, we also collected 328 sets of single mode (sMRI) only subject data in the ADNI database for AD classification performance of the generated model. The preprocessing method for all data is the same.

We used the 18F-FDG-PET (PET data using 18F deoxy glucose) and sMRI data downloaded from ADNI with each pair of FDG-PET and sMRI for the same subject captured at the same time. All sMRI scans (T1-weighted MP-RAGE sequence at 1.5 Tesla) used in our work were acquired from 1.5T scanners and typically consisted of 256 × 256 × 176 voxels with a size of approximately 1 mm × 1 mm × 1.2 mm. The PET images have many different specifications, but they are finally processed into a unified format.

For the sMRI data, we conducted Anterior Commissure (AC) – Posterior Com-missure (PC) reorientation via MIPAV software. Tissue intensity inhomogeneity is then corrected using the N3 algorithm\cite{fenlon2021evolution} . Skull stripping, cerebellum removal, and three main tissues (gray matter (GM), white matter (WM), and cerebrospinal fluid (CSF)) segmentation were conducted via the Cat12 tool of SPM12  . Existing research shows that GM demonstrated higher relatedness to AD\cite{iglesias2021joint,tabari2021comparison} . Therefore, we chose GM masks for this work. Finally, we used the hierarchical attribute matching mechanism for the elastic registration (HAMMER) algorithm\cite{chen2021hierarchical} to spatially register the GM masks to the template of the Montreal Neurological Institute (MNI) 152\cite{he2023quantitative} .

For the existing PET data, we realigned them to the mean image. Then, we regis-tered it to the corresponding sMRI image with the MNI152 brain atlas. Then, all the sMRI and existing PET data were smoothed to a common resolution of 8 mm full width at half minimum. And they were all downsampled to 64 × 64 × 64.

\subsection{Experiment Details}\label{subsec2}
All experiments in this chapter were conducted on a machine with a Linux system and an NVIDIA A100 40G GPU. The algorithm backend was implemented using the PyTorch framework.

The ranges of pixel values of each sMRI or PET are different, hence, we normalized the preprocessed sMRI and PET images to the same range for a subject. We use min-max normalization to scale all pixel values into [0,1] as follows:

\begin{equation}
    z = \frac{x - \min(x)}{\max(x) - \min(x)}
\end{equation}

where z is the normalized pixel value for sMRI or PET.

As previously described, the feature extraction part of the generator consists of three modules. The first two parts are composed of convolution-max-pooling layers and PCA modules, while the last part consists of convolutional layers and residual connec-tions. The image reconstruction part of the generator begins with a self-attention (SA) layer, followed by three deconvolution layers with the same kernel size to upsample the features to the original image size.

During model training, the batch size is set to 16, and the number of training epochs is set to 1000. The learning rates for the generator and discriminator are set to 0.0001 and 0.0004, respectively. Adam optimizer is chosen for both the generator and discriminator, with an exponential decay rate of the first-order moment estimates (beta1) set to 0.9 and the second-order moment estimates (beta2) set to 0.999. For the evaluation of generated images quality, this chapter utilizes three commonly used image evaluation metrics: Mean Absolute Error (MAE), Peak Signal-to-Noise Ratio (PSNR), and Structural Similarity Index (SSIM).

\subsection{Results and Discussion}\label{subsec3}

The cross-modal image generation model proposed in this study achieved an MAE of 0.0194, PSNR of 29.65, and SSIM of 0.9486 for the generated PET images during training on the publicly available ADNI dataset. To validate the effectiveness of the proposed model, we conducted a series of ablation experiments.

\subsection{The impact of the PCA module and SA module}\label{subsec4}

\begin{figure}
    \centering
    \includegraphics[width=1.0\linewidth]{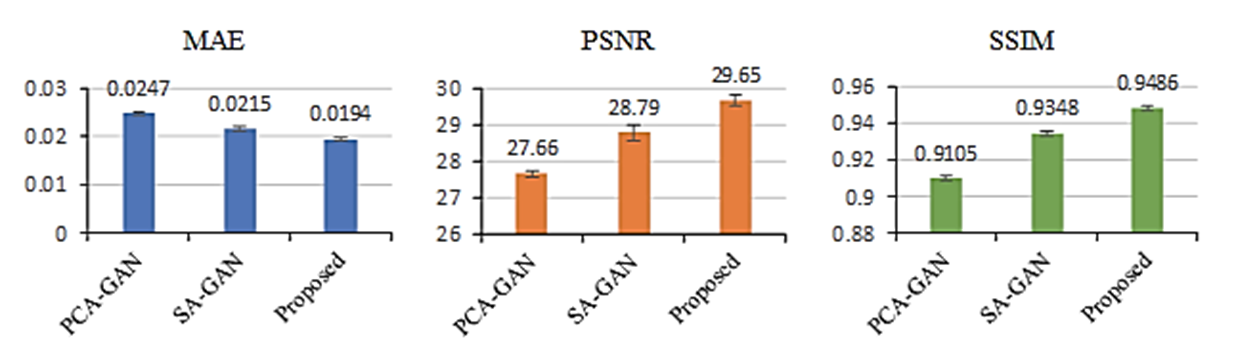}
    \caption{Differences in generated image performances under different methods.}
    \label{fig:Fig3}
\end{figure}

To demonstrate the importance of combining multi-scale local detailed features from the reference image (sMRI) and global correlation information between features during the image generation process, this chapter compares two different architectural variants of the model: (1) PCA-GAN, which removes the self-attention mechanism and only uses the PCA module, and (2) SA-GAN, which removes the PCA module and only uses the self-attention mechanism. The experimental results, as shown in Fig.\ref{fig:Fig3}, validate the effectiveness of combining PCA and SA.

\begin{figure}
    \centering
    \includegraphics[width=0.75\linewidth]{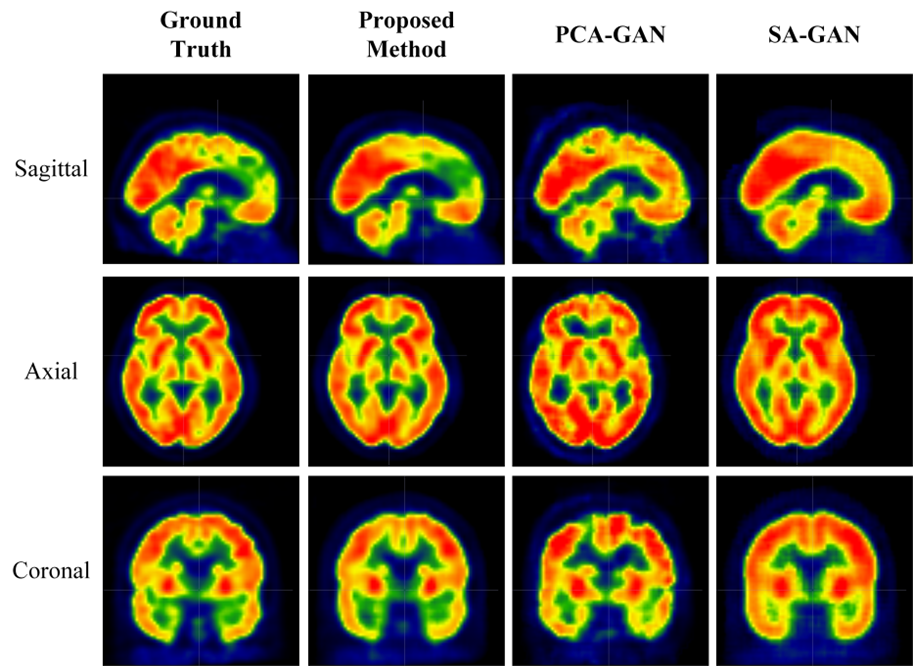}
    \caption{Quality comparisons of generated images (Sagittal, Axial, and Coronal planes).}
    \label{fig:Fig4}
\end{figure}

We also showcased comparisons between real images (Ground Truth), images generated by the proposed model, and images generated by two model variants from three perspectives (sagittal, axial, and coronal planes), as shown in Fig.\ref{fig:Fig4}

Although not very pronounced, there are still some differences in the quality of generated images between PCA-GAN, which is based on multi-scale local features, and SA-GAN, which is based on inter-feature global correlation information. Upon closer observation, it can be noticed that PCA-GAN is better at capturing complex details such as ventricle shape. However, generating images solely based on multi-scale local features may introduce some unnatural details in the overall structure. On the other hand, SA-GAN provides higher-level information that helps maintain overall consistency and coherence in the generated images, but they may lack some fine details, thus reducing the realism. The model that combines both modules achieves better restoration in terms of both local details and global structure.

\subsection{The impact of SA on features of different sizes}\label{subsec5}

SA injects global correlation information into features, thereby ensuring the integrity of the generated images in terms of global structure. The impact of correlation information between features of different sizes on the results may vary. Therefore, we attempted to apply SA at different stages of upsampling (image generation) for different scales of features. The experimental results are shown in Table \ref{tbl:example1}.

The experimental results indicate that the global correlation information between deeper-level features of medical images contributes to generating more realistic PET images.

\subsection{The impact of the loss function}\label{subsec6}
In medical image generation tasks, traditional adversarial loss functions are no longer sufficient, and additional loss functions can help optimize the generator to generate more realistic medical images. Therefore, we conducted experiments to evaluate the impact of different loss functions on the quality of generated images. This involved using the classic adversarial loss function alone and in combination with other loss functions to optimize the parameters of the generator and discriminator models. The experimental results are shown in Table 2.

From the experimental results, it can be observed that additional loss functions can guide the generator to generate higher-quality PET images compared to using only the classical adversarial loss. Furthermore, the joint MM-SSIM loss function shows a more significant improvement in image quality compared to the joint MAE loss function. At the same time, in medical image generation tasks, the multi-scale version of SSIM (MS-SSIM) performs better in optimizing the generator's parameters and generating more realistic PET images than the single-scale SSIM.

\begin{table*}
\centering
\small
  \caption{\ The impact of the self-attention mechanism on generated images for different scales of features.}
  \label{tbl:example1}
  \begin{tabular*}{\textwidth}{@{\extracolsep{\fill}}cccc}
  
    \toprule[1.2pt]
    Patch Size & SSIM($\uparrow$) & PSNR($\uparrow$) & MAE($\downarrow$) \\
    \toprule[0.8pt]
    4×4×4 & 0.9486±0.0011 & 29.65±0.15 & 0.0194±0.0003  \\
    16×16×16 & 0.9346±0.0031 & 28.75±0.21 & 0.0218±0.0011  \\
    32×32×32 & 0.9150±0.0017  & 27.52±0.09  & 0.0251±0.0006 \\
    \toprule[1.2pt]
  \end{tabular*}
\end{table*}

\begin{table*}
\centering
\small
  \caption{\ The impact of the self-attention mechanism on generated images for different scales of features.}
  \label{tbl:example2}
  \begin{tabular*}{\textwidth}{@{\extracolsep{\fill}}cccc}
  
    \toprule[1.2pt]
   Patch Size & SSIM($\uparrow$) &  PSNR($\uparrow$) & MAE($\downarrow$) \\
    \toprule[0.8pt]
    Adversarial Loss & 0.9201±0.0017 & 28.07±0.08 & 0.0235±0.0002  \\
    MAE & 0.9330±0.0013 & 28.68±0.09 & 0.0218±0.0002  \\
    MM-SSIM & 0.9371±0.0027  & 28.92±0.14  & 0.0212±0.0004 \\
    MAE+SSIM & 0.9446±0.0015 & 29.31±0.11 & 0.0201±0.0003  \\
    MAE+ MM-SSIM & 0.9486±0.0011  & 29.65±0.15  & 0.0194±0.0003 \\
    \toprule[1.2pt]
  \end{tabular*}
\end{table*}

\subsection{Comparison with previous studies}\label{subsec7}

\begin{figure}
    \centering
    \includegraphics[width=0.75\linewidth]{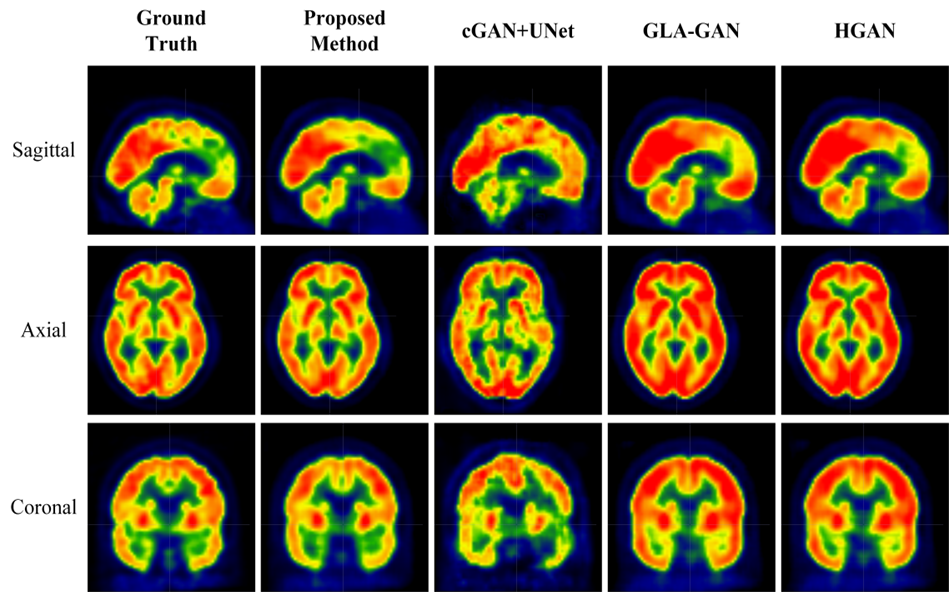}
    \caption{Comparison of image quality generated from previous studies.}
    \label{fig:Fig5}
\end{figure}

To validate the superiority of our proposed method, we first reconstructed a classic cGAN combined with UNet architecture. Additionally, we compared our method with previous state-of-the-art cross-modal image generation models, including HGAN and GLA-GAN, and attempted to reproduce these models as closely as possible based on the model parameter information provided in their respective papers. To ensure a fair comparison, all models were trained using the same dataset as in this chapter. The experimental results are shown in Fig. \ref{fig:Fig5}

The experimental results demonstrate that our proposed method outperforms some of the previous methods in all three image quality evaluation metrics. Fig.\ref{fig:Fig5} also presents the differences in the images generated from the compared methods in the sagittal, axial, and coronal views. The results show that the classical GAN model (3DCNN+UNet) is somewhat limited in handling cross-modal medical image tasks, as the generated PET images appear blurry. In comparison to the Cycle GAN-based HGAN and the patch-based GLA-GAN, our proposed method exhibits superior performance in both fine-texture details and global structures of the generated images, validating its superiority.

\subsection{AD classification performance}\label{subsec8}

In addition to demonstrating the quality of the generated PET images, these generated PET images are used to experiment with their potential to facilitate AD classification. For this purpose, 328 subject data containing only single mode (sMRI) images were collected on the ADNI public dataset and preprocessed accordingly. Then, use the trained cross-modal image generation model to generate corresponding PET images for these data, and input them as a set of multimodal data into the trained classification network model. These data (sMRI and its generated PET) achieved a classification accuracy of 94. 21 \% (AUC: 0.95) on the classification network. Although the classification performance is slightly lower than that of using real data, compared to some previous studies of the same type (sMRI generation PET), the PET images generated by the model proposed in this chapter have better performance in the diagnosis of AD, as shown in Table \ref{tab3}.

\begin{table*}
\centering
\small
  \caption{AD classification performance comparison (\%, mean ± standard deviation)}
  \label{tbl:ad_comparison}
  \begin{tabular*}{\textwidth}{@{\extracolsep{\fill}}lccccc}
    \toprule[1.2pt]
    Methods & ACC($\uparrow$) & SPE($\uparrow$) & SEN($\uparrow$) & F1($\uparrow$) & AUC($\uparrow$) \\
    \toprule[0.8pt]
    Only sMRI & 90.48 & 90.90 & 89.93 & 87.40 & 89.35 \\
    Pan et al.\cite{pan2018synthesizing} & 93.58 & 95.22 & 91.52 & 92.64 & \textbf{96.95} \\
    Sikka et al.\cite{sikka2021mri} & 84.50 & 83.43 & 89.19 & 82.92 & 92.99 \\
    Gao et al.\cite{gao2021task} & 92.00 & 94.00 & 89.10 & 90.50 & 95.60 \\
    Proposed Method & \textbf{94.21} & \textbf{96.39} & \textbf{91.81} & \textbf{93.56} & 94.91 \\
    \toprule[1.2pt]
  \end{tabular*}
\end{table*}

\begin{figure}
    \centering
    \includegraphics[width=0.8\linewidth]{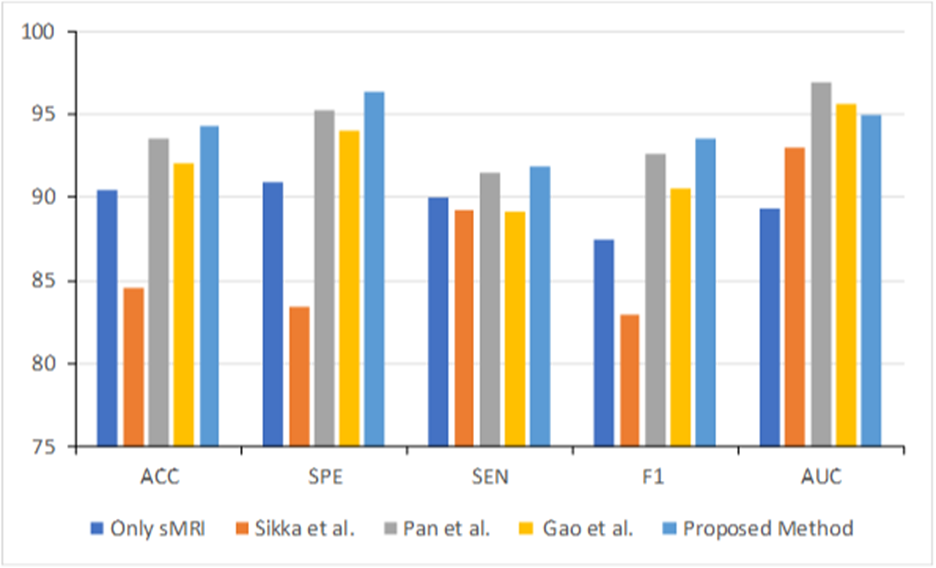}
    \caption{The performance difference in AD classification between single-modality and previous studies.}
    \label{fig:Fig6}
\end{figure}

Furthermore, several parameters were used to evaluate identification performance, including accuracy (ACC), precision, specificity (SPE), recall/sensitivity (SEN), F1 score (F1) and area under receiving operation curve (AUC) (see Fig. \ref{fig:Fig6}). The precision indicates how many of the positive values predicted by the model are positive. F1 is the harmonic average of accuracy and recall/sensitivity, which is a comprehensive evaluation index. AUC can intuitively assess the quality of the identifier.

Since the generation model and classification model in these studies are different from those in this study, it is not easy to implement these methods completely. The proposed method performs better than other methods in AD classification tasks. At the same time, the experimental results also show that the generated PET images can provide more information for the diagnosis of AD (classification performance is higher than with sMRI alone).

\section{Conclusion}\label{sec5}

We propose a cross-modal medical image generation model based on pyramid convolution and attention mechanisms to address the issue of missing information in single-modal (PET) images. To generate more realistic PET images, we utilize pyramid convolution to extract multi-scale feature information from sMRI and employ a channel attention mechanism to alleviate information redundancy in the feature learning process, effectively capturing local detailed features of the image. Before generating the corresponding PET images, the self-attention mechanism is used to incorporate global information among deep-level features, enabling the injection of global dependencies into the multi-scale local features. Additional loss functions are introduced to optimize the parameters of the generator, reducing the gap between generated and real images. The feasibility of the method and the effectiveness of the model are validated through extensive experiments on the ADNI dataset, where the generated images demonstrate good restoration of fine-texture details and global structure. Comparative experiments further confirm the superiority of our model. The experimental results demonstrate that our model generates higher-quality PET images compared to previous research, making it promising for clinical studies.

\bmhead{Acknowledgements}

This work was carried out in part using hardware provided by the High Performance Computing Center of Central South University.





\bibliography{sn-reference}

\end{document}